\documentclass[a4paper,UKenglish]{lipics}

\usepackage{amssymb,amsmath}

\usepackage{microtype}%if unwanted, comment out or use option "draft"
\newcommand{\nameTool}{{\sc \texttt {TouIST}}}

\newcommand{\IMPL}[0]{\longrightarrow}
 
\newcommand{\AND}[0]{\wedge}
\newcommand{\OR}[0]{\vee}
\newcommand{\NOT}[0]{\neg}

\newcommand{\IFF}[0]{\leftrightarrow}
 
\newcommand{\satoulouse}{{\sc Satoulouse}\xspace}

\newcommand{\atL}[2]{{\Large \geqslant}_{#1}^{#2}}
\newcommand{\atM}[2]{{\Large \leqslant}_{#1}^{#2}}
\newcommand{\exact}[2]{{\Large <>}_{#1}^{#2}}
\renewcommand{\satoulouse}{{\sc \texttt {SAToulouse}}}
\renewcommand{\url}[1]{\texttt{#1}}
\title{Twist your logic with \nameTool}
\titlerunning{Twist your logic with \nameTool} %optional, in case that the title is too long; the running title should fit into the top page column

\author[1]{Skander Ben Slimane}
\author[1]{Alexis Comte}
\author[1]{Olivier Gasquet}
\author[1]{Abdelwahab Heba}
\author[1]{Olivier Lezaud}
\author[1]{Frederic Maris}
\author[1]{Mael Valais}

\affil[]{University Paul Sabatier\\
  Toulouse, France\\
  \texttt{\{gasquet,maris\}@irit.fr}}

\authorrunning{S. Ben Slimane et. al.} %mandatory. First: Use abbreviated first/middle names. Second (only in severe cases): Use first author plus 'et. al.'

\Copyright{Skander Ben Slimane, Alexis Comte, Olivier Gasquet\\ 
\phantom{XXXXXXX} Abdelwahab Heba, Olivier Lezaud, Frederic Maris, and Mael Valais}%mandatory, please use full first names. LIPIcs license is "CC-BY";  http://creativecommons.org/licenses/by/3.0/

% mandatory: Please choose ACM 1998 classifications from http://www.acm.org/about/class/ccs98-html . E.g., cite as "F.1.1 Models of Computation". 
% mandatory: Please provide 1-5 keywords
\serieslogo{logo_ttl}%please provide filename (without suffix)
\volumeinfo%(easychair interface)
  {M. Antonia {Huertas}, Jo\~ao {Marcos}, Mar\'ia {Manzano}, Sophie
{Pinchinat}, \\
  Fran\c{c}ois {Schwarzentruber}}% editors
  {5}% number of editors: 1, 2, ....
  {4th International Conference on Tools for Teaching Logic}% event
  {1}% volume
  {1}% issue
  {1}% starting page number
\EventShortName{TTL2015}
\begin{document}

\maketitle

\begin{abstract}
SAT provers are powerful tools for solving real-sized logic problems, but using them requires solid programming knowledge and may be seen w.r.t.\ logic like assembly language w.r.t.\ programming. Something like a high level language was missing to ease various users to take benefit of these tools. \nameTool\ aims at filling this gap. 
It is devoted to propositional logic and its main features are 1) to offer a high-level logic langage for expressing succintly complex formulas (e.g.\ formulas describing Sudoku rules, planification problems,\ldots) and 2) to find models to these formulas by using the adequate powerful prover, which the user has no need to know about. It consists in a friendly interface that offers several syntactic facilities and which is connected with some sufficiently powerful provers allowing to automatically solve big instances of difficult problems (such as time-tables or Sudokus). It can interact with various provers: pure SAT solver but also SMT provers (SAT modulo theories - like linear theory of reals, etc) and thus may also be used by beginners for experiencing with pure propositional problems up to graduate students or even researchers for solving planification problems involving big sets of fluents and numerical constraints on them. 
\end{abstract}


%----------------------------------------------------------------------
\section{The history}\label{sec:introduction}

O. Gasquet and F. Maris teach at University Paul Sabatier in Toulouse, France. They teach logic at different levels starting from introductory courses of propositional logic up to advanced topics for graduate students, like modal logic or logic-based planning. S. Ben Slimane, A. Comte, A. Heba, O. Lezaud and M. Valais are graduate students of the same university. They have been implementing \nameTool\ during three months of their MSc.  

\subsubsection*{Motivation of students}
At the beginning of undergraduate studies, 
we (teachers) found that students' motivation may be increased by showing them that logic is useful and powerful for computer scientists and that computer science does not only consist in hacking C-code or JAVA. 
Classically, logic is motivated by abstract examples or, at the best, by toy examples. At some time, we thought that it would be preferable to show and not only tell them that with little knowledge,
 logic can be used to solve difficult problems whose size prevents humans from 
solving them by hand easily or would require rather complex programming in C or any other programming language. \\

\subsubsection*{\satoulouse's genesis}
In ICTTL'2011, we presented \satoulouse\ \cite{GaScSt2011}, devoted to propositional logic whose main features were 1) to offer a high-level logic language for expressing succinctly complex formulas and 2) to find models of these formulas by using a powerful SAT prover. But \satoulouse\ had several drawbacks to be corrected. \\
Of course, there are loads of logic tools like provers, proof assistants, truth table
editors,\ldots on the Internet, even PROLOG 
 could have been used, but none fits our requirements which are:
\begin{itemize} 
\item the tool must be very easy to install and to use, with no complex syntax; 
 \item the prover can be used as a black box without knowing how it works;
\item no normal forming, ordering on clauses, or PROLOG cut must be needed; 
\item only little knowledge in logic should be necessary.
\end{itemize}

As we could not find an existing tool fulfilling these requirements, in 2010 we
started to implement ours, and we came to the idea of just developing an interface
that allows to very comfortably use a powerful SAT-prover  (namely SAT4J \cite{DBLP:journals/jsat/BerreP10}):
this tool had been called \satoulouse\ and is described in \cite{GaScSt2011}. With this tool, students could
experiment by themselves that a logical language is not only descriptive but may
lead to computations that solve real-life problems. In particular, with
\satoulouse, they could solve Sudokus quite easily, as well as many other
combinatorial problems such as time-table, map coloring, electronic circuits design,\ldots.\

Here are the main facilities that \satoulouse\ offered:
\begin{itemize}
\item Input formulas need not to be in clausal form and arbitrary connectives
  may be used, normal forming is done dynamically during keyboarding of the
  user;
\item Big conjunctions and disjunctions facilities are offered like in:
  \[\bigwedge_{i\in\{1..9\}}
  \bigvee_{j\in\{1..9\}}\bigwedge_{n\in\{1..9\}}\bigwedge_{m\in\{1..9\},m\neq
    n}(p_{i,j,n}\rightarrow \lnot p_{i,j,m})\]
\item Running the solver only consists in clicking a button;
\item The tool displays a model in the syntax of the input formula.
\end{itemize}
Then it is possible to show the power of propositional logic to students that
have been trained a bunch of hours to formalize sentences in logic and have
acquired basic notions of validity and satisfiability to automatically solve some Sudokus.\ 
 
\subsubsection*{Practical work with \satoulouse}
But this
is not the whole story, since the same SAT-solver may be used for solving many
other combinatorial problems as easily as they just did for Sudokus: they just
have to formalize the constraints.\ Our students are asked to do so for:
time-table, map coloring,\ldots \satoulouse{} has been used during three years now by about 400 students with great satisfaction. Particularly, students used it to perform long-term homeworks in the spirit of programming projects: we give them a logical problem to solve (too big to be solved by hand), they must formalize it and then use this formalization to solve the problem. For example, a problem of storage of chemicals that must be stored in same/contiguous/non-contiguous rooms according to their degree of compatibility. Students must solve a case involving a lot of chemicals. 

\subsubsection*{\satoulouse's limitations and \nameTool's genesis}
But during these years, we noticed some painful limitations of \satoulouse: many bugs, flaws in the interface, lack of modularity (if one wishes to change the SAT prover used), ambiguity and limitations of its language, etc.

For example, problems involving pigeon-holes principle like the rules of the Takuzu game\footnote{Also known as Binero. \url{http://fr.wikipedia.org/wiki/Takuzu}} which requires to count 0's and 1's could not be easily formalized: facilities to express something like ``exactly 5 among 10 propositions are true'' were missing. 

\satoulouse\ do not offer the possibility to browse all the models provided by the prover, it only returns one. 

Lessons learned from two years using \satoulouse\ are that many of our CS students clearly become aware that logic has real applications w.r.t.\ problem solving, and many of them gained ability in formalizing problems. But remaining flaws of \satoulouse\ made debugging really hard because only one model is displayed and because of the raw way the models is displayed, together with the poor editing capabilities it has. Moreover only pure combinatorial problems could be handled which heavily limitates the wide range pretention of \satoulouse\ w.r.t.\ real world problems. 

Another drawback of \satoulouse\, not specifically linked to logic teaching, was its inability to be used from the command line: researchers or engineers who wish to use it intensively would find it tedious to type input problems. Last, extension to richer theories is also something that may interest researchers, engineers or graduate students, since \satoulouse\ is definitely not suited for satisfiability modulo theories or for solving planification problems though the same architecture of the software could be used by just changing the solver used. 

A few months ago, we started to go for a whole new software which would fulfill all these demands. It would be called \nameTool\ which stands for TOUlouse  Integrated Satisfiability Tool and should be pronounced ``twist''. 
 
\nameTool\ is of course publicly available for download from the following site 
\begin{center}\url{https://github.com/olzd/touist/releases}\end{center}
%\todo{add address}

To sum it up, here are the features \nameTool\ offers that \satoulouse\ does not:
\begin{itemize}
\item definition of domain sets: $\bigwedge_{i\in A}$ vs. $\bigwedge_{i\in\{Paris,London,Roma,Madrid\}}$
\item multiple binding of indexes: $\bigwedge_{i\in A,j\in B}$ vs. $\bigwedge_{i\in \cdots} \bigwedge_{j\in \cdots}$
\item rich computations on indexes as well as on domain sets $\bigwedge_{i\in (A\cup (B \cap C))}$
\item built-in pigeon-holes primitives: ``atLeast'' (resp. ``atMost'', ``exact'') \emph{so many} values are true among \emph{these values}
\item predicates also may be variables ranging over domain sets: $\bigwedge_{X\in \{A,B\},i\in \{1,2\}} X(i)$ vs. $\bigwedge_{i\in \{1,2\}} (A(i)\AND B(i))$
\item specialized literals targeting constraints between integer or real numbers
\item easy browsing of models successively computed by the solvers
\item regular expressions allowing filtration of literals under interest
\item possibility to use the software on command line and/or batch
\item many editing facilities and improvements
\end{itemize}

%......................................................................
\section{Quick survey of \nameTool}\label{sec:sat_interface}

\nameTool\ is made of three modules, but the standard user will only see one of them: the interface. In the sequel we mainly insist on the latter rather than on the translator and the solver. The global architecture looks as pictured in figure \ref{fig:architectureTouisT}: 

\begin{figure}[htbp]
\includegraphics[scale=0.35]{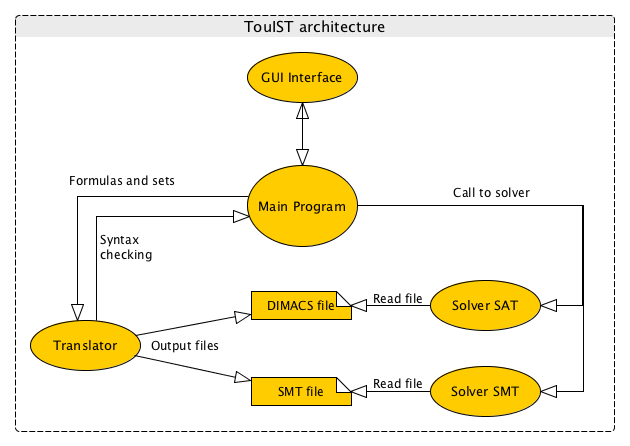}
  \caption{TouIST architecture}
  \label{fig:architectureTouisT}
\end{figure}

With \nameTool\ one accesses a powerful and friendly editor for editing complex logical formulas and various constraints like:

$$\bigwedge_{i \in \{1..9\}} (P_i \IMPL Q_{i+1}),$$

which comfortably abbreviates  $(P_1 \IMPL Q_2) \AND (P_2 \IMPL Q_3) \AND \ldots \AND (P_9\IMPL Q_{10})$. 

\begin{figure}[htbp]
\includegraphics[scale=0.25]{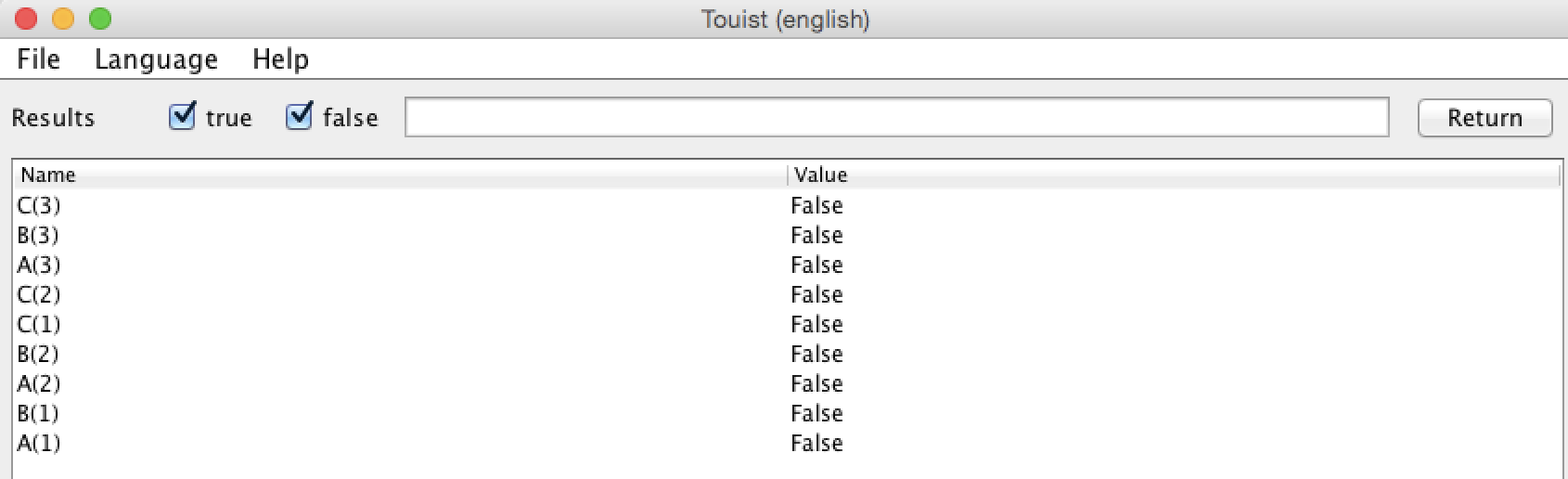}
  \caption{Model display}
  \label{fig:ExampleOfAModel}
\end{figure}

Once it has been given to the interface, a set of formulas may be checked for satisfiability: the interface would send it to the provers which would send back a satisfying model, displayed as shows figure \ref{fig:ExampleOfAModel} if such models exist. Then through the interface, the user can for example ask for other models (button ``Next'' of the interface). 

Models returned by the prover are ``total'' ones: each variable appearing in the formulas sent to the prover is assigned a value. The user may select only True propositions or only False ones. She can also select subsets of the variables under interest by typing a regular expression filtering them. 

\section{Details of what can be done with \nameTool\label{sec:sat_tobedone}}

\subsection{Domain sets}
With time, we noticed that we often need to write things like 
$$\bigwedge_{i \in \{1..9\}} \bigwedge_{j \in \{1..9\}}\bigwedge_{ m\in \{A,B,C,D,E,F,G,H,I\}}P_{i,j,m}\IMPL \bigwedge_{n \in \{A,B,C,D,E,F,G,H,I\}|m\neq n}\NOT P_{i,j,n}$$
If one read $P_{i,j,m}$ as ``there is a letter $m$ in cell $(i,j)$'' of some  $9\times 9$ grid, the above formula expresses that there is \emph{at most} one letter among `A' ... `I' in each cell. 

These sets $\{1..9\}$ and $\{A,B,C,D,E,F,G,H,I\}$ are \emph{domain sets}, with \nameTool\ the user may define as many domain sets she wants, e.g.:

\begin{verbatim}
 N=(1..9)   L=(A,B,C,D,E,F,G,H,I)
\end{verbatim}

and then write the above formula as
$\bigwedge_{i \in N} \bigwedge_{j \in N}\bigwedge_{ m\in L}P_{i,j,m}\IMPL \bigwedge_{n \in L|m\neq n}\NOT P_{i,j,n}$\\
Moreover, usual  operations on sets ($\cup$, $\cap$, $\setminus$, \ldots) can be used to define other sets. 

\subsection{Propositional formulas}

The formulae of \nameTool{} are based on propositional variables (that can have indices) and usual logical operators ($\AND$, $\OR$, $\IMPL$, $\NOT$, $\IFF$). Thus one can type usual simple formulas like $Rain \IMPL Clouds$. But in addition, we provide high-level logical operators that allow to express complex statements in a very compact form. 

\subsubsection*{Generalized conjunctions and disjunctions}
They allow to express conjunctions and disjunctions over formulas containing parameters that vary, e.g.\:
\begin{itemize}
\item $ \bigwedge_{i \in N} P_i$, where $N$ is the domain set defined above. 
  It represents 
  $P_1 \AND P_2 \AND \ldots \AND P_9$. 
\item $\bigvee_{i \in E} P_i$.
\end{itemize}

Of course, these operators may be nested, as in
$\bigwedge_{i \in N} \bigwedge_{j \in N}\bigvee_{ m\in L}P_{i,j,m}$
stating that in each cell there is at least one letter. 

\subsubsection*{Pigeon-hole statements}
They were one of the ``left-to-the-future'' topic of \cite{GaScSt2011}. 
These less classical logical operators are available in \nameTool: they allow to drastically lower the size of some formulas, they are: $\atM{}{}$, $\atL{}{}$ and $\exact{}{}$.\\ The following examples will describe their meanings:
\begin{itemize}
\item $\atM{i \in N}{2} P_i$ represents ``for at most two values of $i \in N$ $P(i)$ is true;
\item $\atL{i \in N}{2} P_i$ represents ``for at least two values of $i \in N$ $P(i)$ is true;
\item $\exact{i \in N}{2} P_i$ represents ``for exactly two values of $i \in N$ $P(i)$ is true;
\end{itemize}
Generalized disjunction is in fact a special case of those: at least one is true, conjuction too: at most 0 are false, and exclusive or may be viewed as: exactly one among two is true. 

Let us recall that with basic logical operators and with $N$ containing 9 elements, $\atM{i \in N}{3} P_i$  would necessitate a formula containing 84 propositions $P_i$ since it amounts to choosing 3 among 9 which yields  $\binom{9}{3}$  possibilities, and neither $\bigwedge$ and $\bigvee$ would help a lot. 

\subsubsection*{Constraints and calculus on indexes}

Often we need to add constraints on indexes, for example:  
$$\bigwedge_{i \in E } \bigwedge_{j \in E  | i \neq j}P_{i,j}$$
which means that $P_{i,j}$ is true whenever $i\neq j$. 

This was the only constraint available in \satoulouse, now in \nameTool\ the range of possibility has been widely enriched. Constraints may include usual comparaison operators like $<$, $>$, $\leq$, $\geq$, $\neq$, $=$ and these comparisons may not only apply to indices but to any arithmetic expressions involving indexes and $+$, $-$, $*$, $/$, $\mod$, $\sqrt{\phantom{x}}$. 
Expressing a sentence like ``each cell $(i,j)$ contains a number which is not equal to $i+j$'' will give:
$$\bigwedge_{i \in N } \bigwedge_{j \in N} \bigvee_{k \in N|k\neq i+j} P_{i,j,k}$$
Of course, \emph{all these sentences} may be expressed with usual plain logical operators, but this would be an aweful work to do. Nevertheless, students must know what is behind the scene, and that such a compact formula abbreviates something long and dull like: 
$$P_{1,1,1}\vee P_{1,1,3}\vee P_{1,1,4}\ldots P_{1,2,1}\vee P_{1,1,2}\vee P_{1,2,4}\vee \ldots $$

\subsection{Technical aspects}

\subsubsection*{Input language vs display language}

\begin{figure}[htbp]
\includegraphics[scale=0.35]{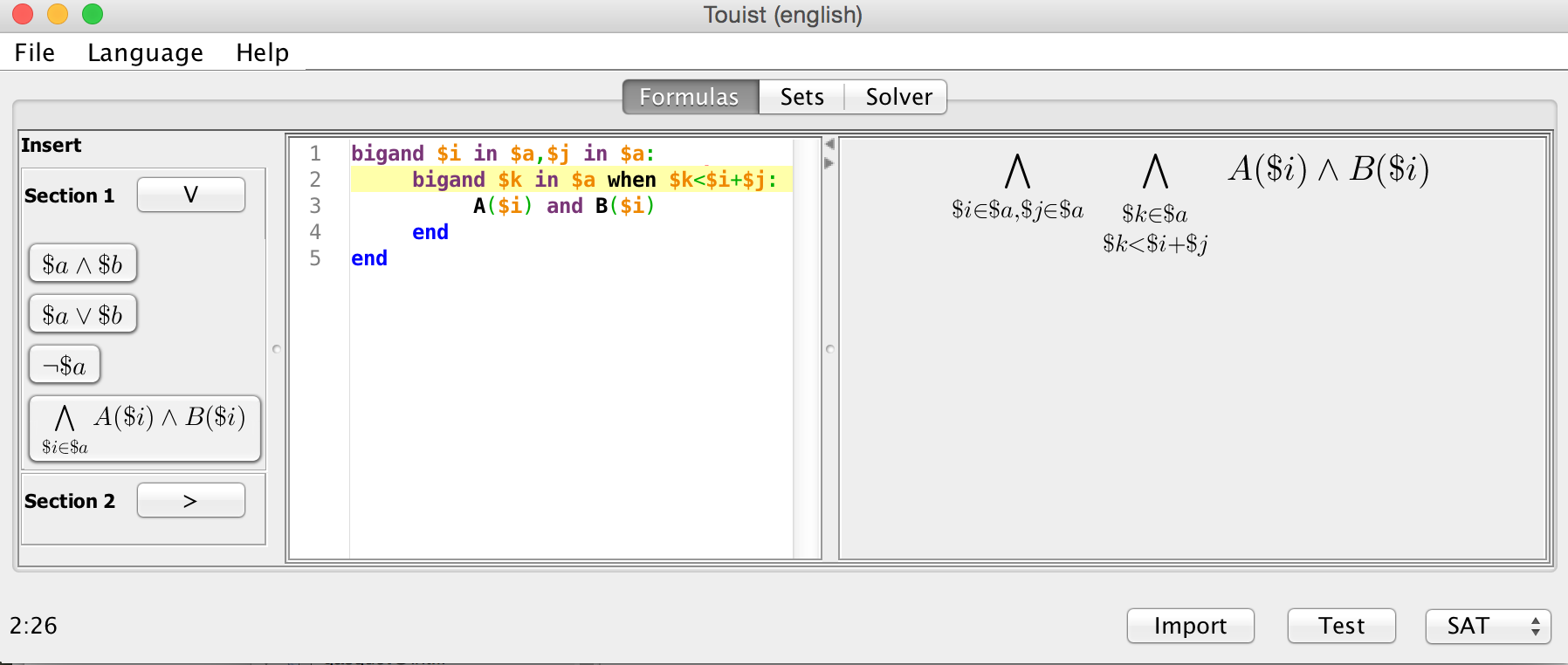}
  \caption{\LaTeX\ style display}
  \label{fig:LatexDisplay}
\end{figure}

Formulas as seen above are written in the \emph{display} language (\LaTeX-style), but all those symbols are not available on keyboards, thus for writing formula and domain sets, the user will use the input language. For example, the above formula together with the associated set $N$ will be typed as (variables are prefixed with \$):
\begin{verbatim}
bigand $i in $N , $j in $N 
  bigor $k in $N when $k < $i+$j :
    P($i,$j,$k))
  end
end
\end{verbatim}
But \nameTool\  displays it in \LaTeX-style as seen in the right panel shown in figure \ref{fig:LatexDisplay}. The definition of the set $N$ is done in the Sets tab. 

Also, formulas may either be
hand-typed in the editor window, or introduced in a sort of syntax-directed editor, by progressively refining the syntax tree, or else they can be imported from some external file.

\section{Advanced topics for graduate students \label{sec:advanced_topics}}

In what follows, we very briefly present some advanced features of \nameTool. They may rather interest researchers, engineers, graduate students and their teachers. They concern SMT (SAT modulo theories), Planning as SAT and their combination Planning as SMT. 

\subsection{SMT: SAT modulo theories}

Some combinatorial problems require nevertheless to deal with some calculus over natural or real numbers. This can be done using only propositional logic (e.g.\ $2+3=5$ may be encoded by $ADD_{2,3,5}$), but it is very uncomfortable as soon as there are more than a few additions to be made. Do not even mention products or more complex operations. The idea behind SMT genesis has been to combine SAT solvers with arithmetic solver in order to improve the treatment made to the arithmetic part of reasoning. In many cases, it will not only improve the efficiency of the prover, but will also allow to express arithmetic constraints of problems in a drastically more compact way. 

Think of the Kamaji game\footnote{\texttt{http://fr.wikipedia.org/wiki/Kamaji}}  where the player must group adjacent numbers in a grid so that their sum is equal to some fixed number. Solving the game essentially requires logical reasoning but still needs a few arithmetic (addition). 

Then if $x_{i,j}$ for each cell $(i,j)$ is an integer and $G(i,j,i,k)$ represents the fact that cells $(i,j)$ to $(i,k)$ of line $i$ form a group, the sum constraint may be expressed as: $\Sigma_{m\in E}x_{i,m}=N$ 
where $N$ is the fixed number and $E$ is $\{j,j+1,\ldots,k\}$. Pure propositional logic is definitely unsuited for such sentences! 

\subsection{\nameTool\ for classical planning as SAT}

In Artificial Intelligence, \emph{planning} is a cognitive process to automatically generate, through a formal procedure, an articulated result in the form of an integrated decision-making system called \emph{plan}. The plan is generally in the form of an organized collection of \emph{actions} and it must allow the universe to evolve from the \emph{initial state} to a satisfactory state, the \emph{goal}.
Propositional planning as SAT has been introduced by Kautz and Selman in  \cite{kautzS92_planning_sat}.

One important difference of \nameTool\, compared with \satoulouse\, is its ability to take into account both logic formulas and domain sets. For example, if one wants to solve a particular planning problem, \satoulouse\ is easy to use for describing the problem and solving it via a SAT solver. But in order to solve several generic planning problems, we can take advantage from the flexibility of \nameTool\ which will allow the user to describe a generic solving method with rules encoded as formulas and to use domains sets to describe each particular planning problem. Numerous encoding rules for planning problem resolution have already been proposed  \cite{kautzS92_planning_sat,MaliK99_plan_space_encodings,Rintanen:2006}. As an example of such a rule we give below an encoding of frame-axioms. If a fact is false at step i-1 of a solution plan and becomes true at step i, then the disjunction of actions that can establish the fact at step i of the plan is true. That is, at least one of the actions that can establishes the fact should have been applied.

  \[\bigwedge_{i\in\{1..length\}}
  \bigwedge_{f\in \mathit{Facts}}\left((\lnot f(i-1) \wedge f(i)) \Rightarrow 
  \bigvee_{a\in \mathit{Actions} / f\in \mathit{Effects}(a)} a(i)\right)\]

\subsection{\nameTool\ for temporal planning as SAT (modulo unquantified rational difference logic)}

Moreover, in addition to SAT, our new platform \nameTool\ is able to handle theories like difference logic or linear arithmetic on integer or real numbers, and call a SMT solver to find a solution. To be solved, real world temporal planning problems require to represent continuous time, and so, the use of real numbers in logic encodings. \nameTool\ can also be used to solve such problems involving durative actions, exogenous events and temporally extended goals, for example with encoding rules proposed in \cite{MarisRegnier08}. We give below an encoding of temporal mutual exclusion of actions. If two actions respectively producing a proposition $p$ and its negation are active in the plan, then the time interval $[\tau(a\mid\rightarrow p),\tau(a\rightarrow\mid p)]$ corresponding to the activation of $p$, and the time interval $[\tau(b\mid\rightarrow\neg p),\tau(b\rightarrow\mid\neg p)]$ corresponding to the activation of $\neg p$ are disjoint.
  \[\bigwedge_{a\in Actions}
  \bigwedge_{b\in Actions}
  \bigwedge_{f\in Facts | f\in \mathit{Effects}(a) \wedge \lnot f\in \mathit{Effects}(b)}\]
  \[\left[\left(a \wedge b\right) \Rightarrow 
  \left[ \left(\tau(b \rightarrow\mid \lnot f) < \tau(a \mid\rightarrow f)\right)
    \vee
    \left(\tau(a \rightarrow\mid f) < \tau(b \mid\rightarrow \lnot f)\right)
 \right]\right] \]

%----------------------------------------------------------------------
\section{Conclusion}\label{sec:evaluation}

As far as we are aware, there is no other tool targeted at the same large audience, neither at the same wide class of problems, neither with the same comfort.  Most existing
pedagogical tools (either implementation of truth-tables or semantic tableaux)
that could do the job of searching a model cannot efficiently handle big
problems, and real tools able to deal with them are definitely not designed to
be used by beginners in logic, and not even by most graduate students. Advanced tools designed for graduate topics, like Mozart \cite{DBLP:conf/moz/2004} or Alloy  \cite{Jackson:2006:SAL:1146359} have a steep learning curve that may dissuade beginners and non-specialist users. 

We believe \nameTool\ will be useful for beginners in logic as well as for advanced users thanks to its large scope of applications and to its ease of use. 

%----------------------------------------------------------------------
\bibliographystyle{plain}

\begin{thebibliography}{}

\end{thebibliography}


\begin{thebibliography}{50}

\bibitem{DBLP:journals/jsat/BerreP10}
Daniel~Le Berre and Anne Parrain.
\newblock The sat4j library, release 2.2.
\newblock {\em JSAT}, 7(2-3):59--6, 2010.

\bibitem{GaScSt2011}
Olivier Gasquet, François Schwarzentruber, and Martin Strecker.
\newblock Satoulouse: the computational power of propositional logic shown to
  beginners.
\newblock In P.~Blackburn, H.~van Ditmarsch, M.~Manzano, and F.~Soler-Tosca,
  editors, {\em Third International Congress on Tools for Teaching Logic
  (ICTTL'2011)}, volume 6680 of {\em Lecture Notes in Computer Science}, pages
  77--84. Springer, 2011.

\bibitem{Jackson:2006:SAL:1146359}
Daniel Jackson.
\newblock {\em Software Abstractions: Logic, Language, and Analysis}.
\newblock The MIT Press, 2006.

\bibitem{kautzS92_planning_sat}
Henry Kautz and Bart Selman.
\newblock Planning as satisfiability.
\newblock In {\em IN ECAI-92}, pages 359--363. Wiley, 1992.

\bibitem{MaliK99_plan_space_encodings}
Amol~Dattatraya Mali and Subbarao Kambhampati.
\newblock On the utility of plan-space (causal) encodings.
\newblock In {\em Proceedings of the Sixteenth National Conference on
  Artificial Intelligence and Eleventh Conference on Innovative Applications of
  Artificial Intelligence, 1999}, pages 557--563, 1999.

\bibitem{MarisRegnier08}
Fr\'ed\'eric Maris and Pierre R\'egnier.
\newblock Tlp-gp: New results on temporally-expressive planning benchmarks.
\newblock In {\em International Conference on Tools with Artificial
  Intelligence (ICTAI)}, volume~1, pages 507--514. IEEE Computer Society, 2008.

\bibitem{Rintanen:2006}
Jussi Rintanen, Keijo Heljanko, and Ilkka Niemel\"{a}.
\newblock Planning as satisfiability: Parallel plans and algorithms for plan
  search.
\newblock {\em Artif. Intell.}, 170(12):1031--1080, 2006.

\bibitem{DBLP:conf/moz/2004}
Peter~Van Roy, editor.
\newblock {\em Multiparadigm Programming in Mozart/Oz, Second International
  Conference, {MOZ} 2004, Charleroi, Belgium, October 7-8, 2004, Revised
  Selected and Invited Papers}, volume 3389 of {\em Lecture Notes in Computer
  Science}. Springer, 2005.

\end{thebibliography}

\end{document}